\documentclass{article}
\usepackage{graphicx}
\usepackage[round]{natbib}
\usepackage{epsfig}
\usepackage{amsmath}
\title{
Change-point detection in the historical hurricane number
time-series: why can't we detect change-points at US landfall?
}

\author{
Kechi Nzerem (RMS)\\
Stephen Jewson (RMS)\footnote{\emph{Correspondence email}: \texttt{stephen.jewson@rms.com}}\\
Thomas Laepple (AWI)\\
}

\begin{document}
\maketitle

\begin{abstract}
The time series of the number of hurricanes per year in the Atlantic basin
shows a clear change of level between 1994 and 1995. The time
series of the number of hurricanes that make landfall in the US,
however, does not show the same obvious change of level. Prima-facie this
seems rather surprising, given that the landfalling hurricanes are a subset
of the basin hurricanes.
We investigate whether it really should
be considered surprising or whether there is a simple statistical
explanation for the disappearance of this change-point at landfall.
\end{abstract}

\section{Introduction}

Recent increases in both the number of hurricanes in the Atlantic
and the number of hurricanes making landfall on the US coastline
motivate a desire to try and understand whether the frequency distribution of
hurricane occurrence varies in time, and in particular whether it has changed recently.
These questions have been addressed
by a number of previous authors and in general there is agreement
that the distribution of the number of hurricanes per year in the
Atlantic basin is not stationary, but has shown changes throughout
the last 100 years, including recent changes (see, for example,
\citet{elsnerj00}, \citet{elsnern03}, \citet{goldenberg}).
There is no clear agreement about what exactly is
causing these changes, but most authors seems to believe that they
are driven by a combination of climate change and internal climate
variability.

One of the statistical methods that has been used to understand
the historical behaviour of hurricane numbers in the Atlantic basin is
change-point analysis, in which one tries to identify points in
time at which the mean of the time-series changed. Such a
change-point analysis of hurricane numbers was performed
by~\citet{elsnerj00} and~\citet{elsnern03}, who analysed category 3-5 hurricane numbers
in the Atlantic basin and found change-points in the years 1942/1943, 1964/1965
and 1994/1995. \citet{e02a} did a further analysis of both category 1-5
and category 3-5 hurricane numbers in the basin
(using a very different statistical method from the methods used in
\citet{elsnerj00} or~\citet{elsnern03}) and found change-points in the years
1931/1932, 1947/1948, 1969/1970 and 1994/1995 for category 1-5 storms
and the years 1914/1915, 1947/1948, 1964/1965 and 1994/1995
for category 3-5 storms.
Comparing the results from these different studies we see that the 1994/1995
change is the most robust, since it appears in all the studies.
It is also the most important since it is
the most recent, and hence the most relevant to understanding the
current and future levels of hurricane activity. One should be a
little cautious about interpreting the earlier change-points as
necessarily telling us something about physical reality, since the
earlier data is possibly somewhat inaccurate.

%One might, for instance, make a prediction of the next few years by assuming
%that there will be no further change-points. Such an assumption seems reasonable for the near term
%future, but gradually becomes less reasonable the further into the future that one wants to predict.

\citet{e02b} repeated their change-point analysis for the number
of hurricanes at US landfall. The results were strikingly
different from the basin change-point results: they couldn't
identify a single significant change-point in either the series of
category 1-5 storms or the series of category 3-5 storms.
These results agreed with the earlier work of~\citet{elsnern03}
who also couldn't identify change-points in the landfalling series.
Prima
facie, these results seem rather surprising. The landfalling storms, are,
after all, a subset of the total number of storms, and might
naively be expected to inherit the properties of the basin time
series. Why, then, do the change-points disappear, and why, in
particular, does the 1994/1995 change-point disappear? One can
imagine two possible answers to this question: (a) that the
process by which basin storms become landfalling storms is
complex, and varies in time in such a way that the landfalling
series does not inherit properties of the basin series, or (b)
that the process by which basin storms become landfalling storms
is simple, and dominated by the geometry of typical hurricane
tracks and the shape of the US coastline, and the landfalling
series does in principle inherit the properties of the basin
series, but that in practice these properties are obscured by
noise in the landfalling series. Reality may be
a combination of these two limiting-case explanations.

One first test, which sheds some light on these two hypotheses,
is to assess whether hypothesis (b) is a plausible one or not, and
that is the goal of this article. We will do this
by considering only the 1994/1995 change-point. We look at how easy it is to detect this
change-point in both the basin and the landfall data, and then we look at how easy one would \emph{expect}
it to be to detect the change-point in both the basin and the landfall data,
given that there is a change-point in the basin
data, and given a simple relationship between basin and landfall.
By doing so we are investigating whether the `disappearance'
of the change-point at landfall in the real data could be just a simple statistical effect.

\section{Results}

\begin{figure}[!hb]
  \begin{center}
    \rotatebox{-90}{\scalebox{0.7}{\includegraphics{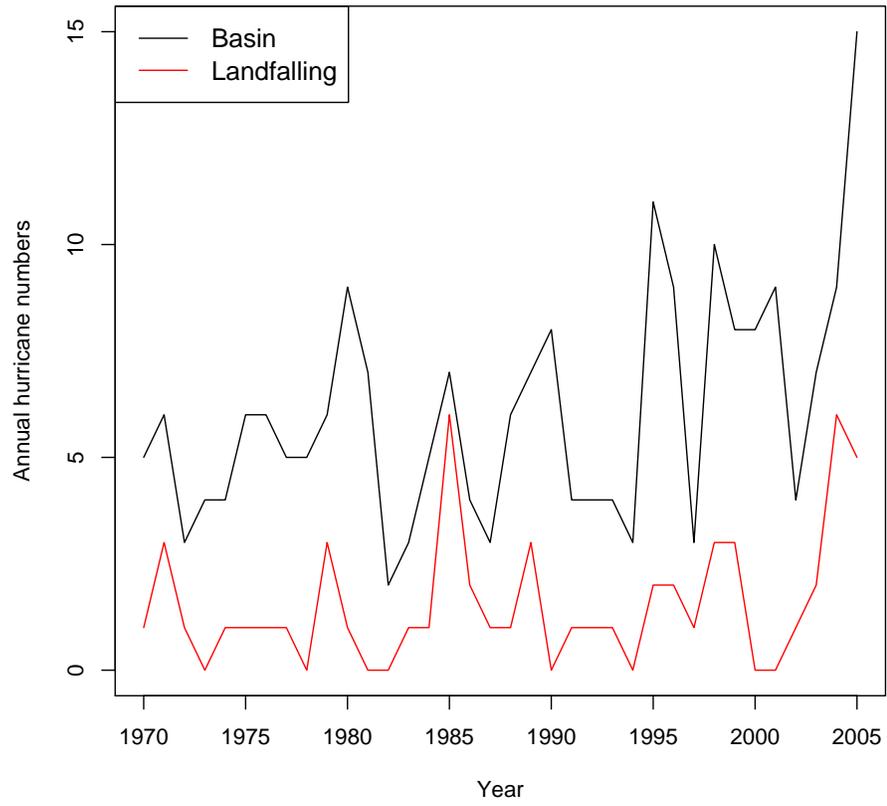}}}
  \end{center}
    \caption{
Atlantic basin and landfalling hurricane numbers for the period 1970 to 2005.
}
     \label{f01}
\end{figure}

For the observed basin hurricane numbers, derived from HURDAT~\citep{hurdat}
and shown in figure~\ref{f01} (black line),
we first test whether there is a significant difference in
the mean number of hurricanes between the periods 1970-1994 and
1995-2005. These periods are defined by the last two change-points
detected by~\citet{e02a} in the total hurricane number time-series.
We use a statistical test
that compares the differences in the means of poisson
distributions for finite samples (see the appendix for more
details: this test, known as a `C-test', is a poisson distribution
version of the well-known t-test that can be used to detect
differences in the means of data from a normal distribution). We
would expect to find a significant difference, since the 1994-1995
change-point was identified using a statistical method
based on the same data (in other words, this is not really a well-posed test
since we are using the data twice: once to create the hypothesis, and once to test it.)
We obtain a p-value for this
difference of 0.000258, which is indeed highly significant.
However, when we repeat the same test for
landfalling hurricane numbers (also shown in figure~\ref{f01}, grey line),
we only obtain a p-value of 0.0364 (and this is now a well-posed test).  We
would thus reject at the 0.01 level the hypothesis that the rates
before and after the change-point are different for landfalling
hurricanes, but we would accept it for basin hurricanes.

We now address the question of \emph{why} we might be seeing this difference in significance
levels between the basin and landfall data using simulations.

\begin{figure}[!hb]
  \begin{center}
    \rotatebox{-90}{\scalebox{0.7}{\includegraphics{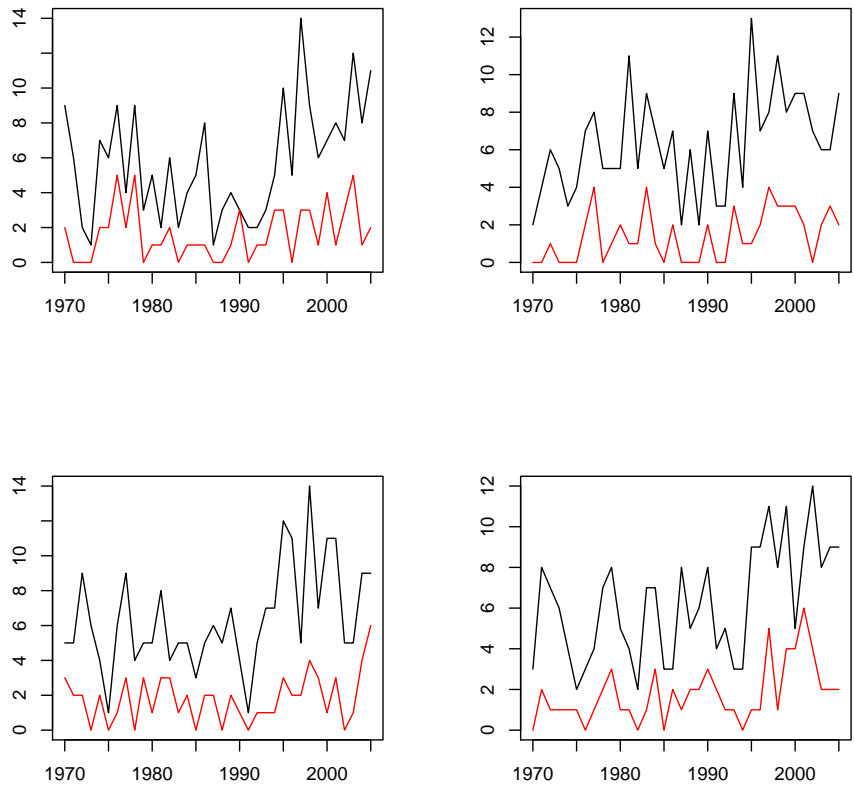}}}
  \end{center}
    \caption{
Four realisations of simulated Atlantic basin and landfalling hurricane numbers for the period 1970 to 2005,
with an artificial change-point included in both series between 1994 and 1995.
}
     \label{f02}
\end{figure}

First, we generate 10,000 realisations of surrogate data for the number of basin hurricanes for the
period 1970 to 2005 by simulating from independent poisson distributions.
We incorporate an artificial change-point into this data between 1994 and 1995, by setting
the mean from 1970-1994 at 5.04 (which is the historical mean number of basin hurricanes for this period)
and the mean from 1995-2005 at 8.45 (likewise).
4 of these 10,000 realisations are shown in figure~\ref{f02} (black lines).
We then test whether we can detect the change-point
by analysing each 36-year surrogate series individually.
At the p=0.01 level we find that we detect the change-point in 8,427 of our 10,000 cases
(i.e. 84\% of the time). In the rest of the cases the change-point is obscured by the noise
and we can't detect it even though we know it's there because we put it there ourselves.

Second, we derive 10,000 realisations of surrogate data for the number of \emph{landfalling} hurricanes
for the same period.
We derive the landfalling numbers from the simulated basin data by applying probabilities that each basin simulated hurricane will make landfall.
The probabilities we use are estimated separately from the observations for the two periods prior to and post 1994/1995, and
are $0.246$ and $0.269$.
Combining the mean hurricane numbers for the basin with the probabilities of making landfall,
we see that the mean numbers of landfalling
hurricanes on either side of the change-point (in reality and in our simulations) are 1.24 and 2.27.
The fact that there is a slightly higher probability of making landfall in the second
period will increase the average number of landfalling hurricanes in the second period, relative to a situation
in which we had used a single probability estimated on data from 1970-2005.
This will make it
easier to detect the landfalling change-point.
4 of the 10,000 realisations of landfalling hurricane numbers are shown in figure~\ref{f02} (grey lines), alongside
the corresponding basin simulation.
Once again we test each of our 10,000 realisations to see
if we can detect the change-point. However, in this case we only detect the change-point in 3660 of our 10,000 tests
(i.e. 37\%) of the time.
In the rest of the cases we are again in the situation where we can't detect it even though we know it's there.

In other words, even though there is a large change-point built-in to the surrogate landfall data
(corresponding to an increase of 83\% in the mean)
it is hidden by noise in most of our realisations.

Finally, we compare the distribution of p-values derived from the statistical tests on the
simulations with the p-values from the statistical tests on the
historical data.
The basin p-value for the real data is at the 49th percentile of the simulated distribution of basin p-values,
and the landfall p-value for the real data is at the 46th percentile of the simulated distribution of the landfalling p-values.
We conclude that both the historical p-values are entirely consistent with the simulations.

\section{Conclusions}

\begin{figure}[!hb]
  \begin{center}
    \rotatebox{-90}{\scalebox{0.7}{\includegraphics{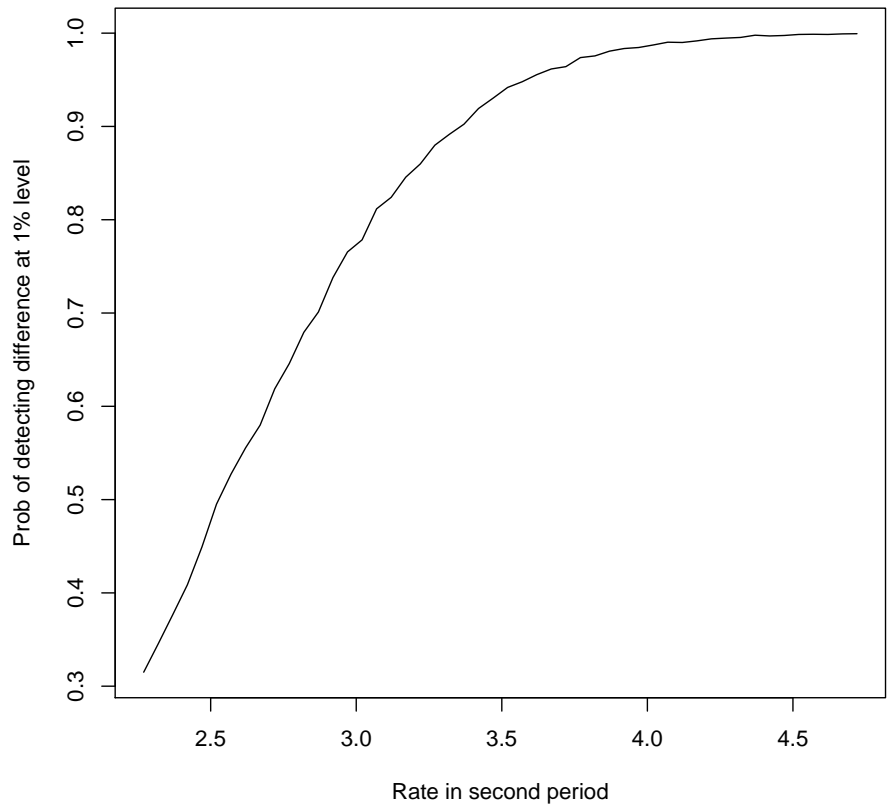}}}
  \end{center}
    \caption{
The probability of detecting the 1994/1995 change-point versus the underlying landfalling hurricane
rate from 1995 to 2005.}
     \label{f03}
\end{figure}

We have asked the question: why is it possible to detect change-points in the time series of the number
of hurricanes in the Atlantic basin, but not possible to detect change-points in the time series of the number
of hurricanes at US landfall? To simplify matters we have considered only the most recent change-point,
which occurred between
the years 1994 and 1995. Using Monte-Carlo simulations we have shown that although one would expect
to be able to detect a change-point of the same magnitude as the observed 1994/1995 change-point most
of the time in the basin data, one would \emph{not} expect to be able to detect the impact of this change-point
in the landfall data, for the simple
statistical reason that the change disappears in the noise. We thus conclude that, although there may be
other things going on that also conspire to hide the change-points in the landfalling time-series, one doesn't
\emph{need} to invoke anything other than simple statistics to explain the apparent lack of change-points in this series.
Or, to put it another way, one cannot conclude from the lack of detectable change-points in the landfall series
that this series isn't changing. Quite significant changes could have occurred, and yet still not be
detectable, because of the low number of hurricanes making landfall, and the implied signal to noise ratio.
Figure~\ref{f03} illustrates this by showing the probability of detecting the 1994/1995 change-point
in the landfalling series, versus the level of hurricane activity from 1995-2005. To reach a 50\%
chance of detecting a change-point in the landfalling data,
the annual landfalling hurricane rate after 1995 would have had to have been over
2.5 hurricanes per year (over a 100\% increase relative to 1970-1994). To reach a 90\% chance of detecting a change-point the average landfalling hurricane rate
after 1995 would have to have been at around 3.5 hurricanes per year (nearly a 200\% increase).

Overall the disappearance of change-points at landfall occurs because when we reduce the number of hurricanes by a factor of 4 (as we do going from basin
to landfall) the size of any change-points reduces by a factor of 4, but the standard deviation of the poisson distribution
only reduces by a factor of 2 (this is a property of the poisson distribution). The signal-to-noise ratio thus
decreases by a factor of 2, making change-points twice as hard to detect in the landfalling data as they are
in the basin data.

\section{Appendix}

We now describe the details of our statistical test, which is due to~\citet{przy}.

The number of basin hurricanes in year t, N$_{B,t}$, is assumed to
follow a poisson distribution with rate $\lambda$.
Conditional on a given number of basin hurricanes
N$_{B,t}$ = n$_{B,t}$ we then model the number
making landfall N$_{L,t}$ as
following a binomial distribution with probability p, so that
\[N_{L,t} \sim Bin(n_{B,t},p)\]

It is then easy to show that the number at landfall is then also given by
a poisson distribution with rate $\lambda$p.

We assume that there is a change in the value of $\lambda$ in
1994/1995 and that it takes one constant value $\lambda_1$ between
1970 and 1994, and then another constant value $\lambda_2$ between 1995 and 2005.

We estimate $\lambda_i$ by the sample mean $\frac{1}{n_i}
\sum_{j=1}^{n_i}N_{B_i,j}$, and we estimate the conversion rate p$_i$ of basin to
landfalling hurricanes in period i to be the proportion
N$_L{_i}$/N$_B{_i}$ where N$_L{_i}$ and N$_B{_i}$ are the numbers of
landfalling and basin hurricanes in period i.

\subsection{Hypothesis Tests}

We consider the two variables

\begin{equation}
B_{tot,i}=\sum_{t=1}^{n_i}, \hspace{3mm} L_{tot,i}=\sum_{t=1}^{n_i}, i=1,2
\end{equation}

which are the total numbers of basin and
landfalling hurricanes in period i.
These variables are
themselves also poisson variables with

\begin{equation}
B_{tot,i} \sim Pois(n_i\lambda_i), \hspace{3mm} L_{tot,i} \sim Pois(n_i\lambda_ip)
\end{equation}

We wish to
test the hypothesis

\begin{equation}
H_0: \lambda_1 = \lambda_2 \hspace{3mm} vs \hspace{3mm} H_a: \lambda_1 \neq \lambda_2
\end{equation}

The conditional test (C-test) (due to Przyborowski and Wilenski
(1940)) is based on the fact that the conditional distribution of
$B_{tot,1}$ given $B_{tot,1}+B_{tot,2}=k$ is binomial
\[P(B_{tot,1}=k_1) = \binom{k}{k_1}p^{k_1}(1-p)^{k-k_1}\] with probability
\[p=p(\lambda_1/\lambda_2) =
(n_1\lambda_1)/(n_2\lambda_2)/(1+(n_1\lambda_1)/(n_2\lambda_2))\]

We reject $H_0$ at the $\alpha \%$ level if

\begin{equation}
2min\lbrace P(B_{tot,1} \ge k_1|k,p(1)), \hspace{3mm} P(B_{tot,1} \le k_1|k,p(1))\rbrace \ge \alpha
\end{equation}

Here we wish to test for the ratio $\lambda_1/\lambda_2=1$, so
$p=p(1)$.

We test the numbers of landfalling hurricanes in the same way.

\bibliography{arxiv}

\newpage

\newpage

\newpage

\end{document}